# Lateral vibration of footbridges under crowd-loading: Continuous crowd modeling approach

Joanna Bodgi[1, a], Silvano Erlicher[1,b] and Pierre Argoul[1,c]

Institut NAVIER, ENPC, 6 et 8 av. B. Pascal, Cité Descartes, Champs sur Marne,

77455 Marne-la-Vallée, France

[a] bodgi@lami.enpc.fr, [b] erlicher@lami.enpc.fr, [c] argoul@lami.enpc.fr

**Keywords:** Footbridges, Crowd loading, Pedestrian-structure synchronization.

**Abstract.** In this paper, a simple 1D crowd model is proposed, which aim is to properly describe the crowd-flow phenomena occurring when pedestrians walk on a flexible footbridge. The crowd is assumed to behave like a continuous compressible fluid and the pedestrian flow is modeled in a 1-D framework using the (total) mass (of pedestrians) conservation equation. This crowd model is then coupled with a simple model for the dynamical behavior of the footbridge and an optimized modeling of synchronization effects is performed. Numerical simulations are presented to show some preliminary results.

**Introduction**

Recent examples of footbridges have shown to be sensitive to the human induced vibration (Millenium Bridge, London; Solférino Bridge, Paris). Several experimental measurements allowed this phenomenon to be better understood [1, 2]. The crowd walking on a footbridge imposes to the structure a dynamic lateral excitation at a frequency close to 1 Hz. When the first mode of lateral vibration of a footbridge falls in the same frequency interval, then a resonance phenomenon is activated, the oscillation amplitude increases and pedestrians are forced to change their way of walking, up to the so-called structure-pedestrian synchronization occurring if the oscillation amplitude is large enough. This phenomenon has been often experimentally detected and also analyzed in several studies [1, 3, 4]. A series of simplified design rules for footbridges accounting for these effects was recently proposed [2].

Moreover, the behaviour of a single pedestrian is affected by the presence of the crowd around him. In more detail, when the pedestrian density is very low, the walk is « free » and characterized by the speed, the walk frequency, the step length, etc. slightly varying from a walker to the other. Nonetheless, when the crowd density becomes higher, a single walker is forced to synchronize his speed with that of the others. This kind of pedestrian-pedestrian synchronization occurs even when the walk is on a rigid floor and some crowd models were already developed [5] for this case. However, very few existing studies concern the modeling of both kinds of synchronizations [3]. In this contribution, an approach that we call "Eulerian", is proposed in order to take into account traffic effects and pedestrian-structure synchronization.

**Some experimental results**

**Single pedestrian's walking.** The human walking is characterized by time intervals where both feet are in contact with the floor and intervals where only one foot touches the floor. One can define the beginning of a step as the beginning of a simultaneous contact period. The end of a step coincides with the beginning of the following simultaneous contact period. For a given pedestrian walking at constant speed, the time-length $T_s$ of a step is approximately constant. The walking frequency is defined by: $f_s = 1/T_s$. For a standard walking, one gets $f_s \approx 2 Hz$ [4]. The force induced by a single pedestrian on the floor has a lateral component related to the small lateral oscillation of the centre of gravity of the pedestrian during walk. It acts in the direction perpendicular to the walk speed, and with opposite signs for each foot. Hence, the frequency of the lateral force is: $f_{lat} = f_s/2 \approx 1 Hz$. The processing of data provided by Decathlon [6] concerning lateral walking forces show a quasi-periodic behavior, with a typical right-left cycle like the one indicated in Fig. 1. The periodicity of the lateral force $f_l(t)$ suggests a Fourier harmonic decomposition:

$$f_l(t) = \sum_{i=1}^{+\infty} \alpha_i \sin(2\pi i f_{lat} t - \varphi_i) \approx \alpha_1 \sin(2\pi f_{lat} t - \varphi_1) \qquad (1)$$

where $\alpha_i$ is the $i$-th Fourier coefficient, $\varphi_i$ is the $i$-th phase difference. $f_l(t)$ is always bounded and less than $50 N$. For an increasing walking speed, the frequency $f_{lat}$ also increases.

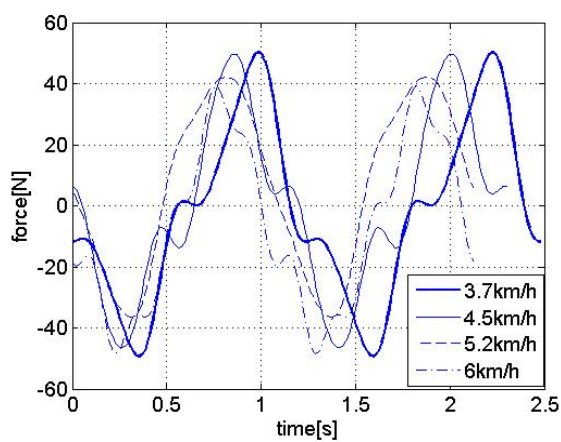

Figure 1- Lateral walking force on a rigid floor, for several walking speeds.

It can be deduced that the first harmonic amplitude is of order $30-35 N$, confirming the results of SETRA [2] and can be related to the walking velocity $v$ by: $\alpha_1 = 0.6191 v + 35.5171$. For the analysis of the crowd effect on the lateral motion of footbridges, only the first harmonic (frequency $f_{lat}$) is retained in most studies and this will be the case in this paper. This $f_{lat}$ frequency is often very close to the lateral modal frequency of footbridges. Hence footbridges may be subject to large oscillations because their behavior is often slightly damped even if the pedestrian lateral force is relatively small. In the case of non-rigid floor, the lateral force amplitude of the human walk is assumed to be similar to that of the rigid-floor case. This strong assumption is supported by the experimental results of SETRA [2]. Conversely, the walking velocity and frequency are affected by the structure's vibrations, as seen in the following paragraph.

**Influence of the structural oscillation on pedestrians walking.** When the amplitude of the lateral vibrations acceleration for the footbridge is larger than a certain threshold: $a_{min} = 0.1 m/s^2$, the lateral vibrations become perceptible for pedestrians, who tend to change their walking frequency to synchronize their walk with the structure's oscillations [4]. If a certain number of pedestrians is synchronized with the structure (and therefore, with each other), the total lateral force they produce further increases the structure's oscillations, inducing other pedestrians to synchronize their walk. If the lateral velocity of the footbridge floor reaches $\dot{u}_{max} = 0.25 m/s$, pedestrians stop walking, otherwise they would loose equilibrium [3].

**Interactions between pedestrians.** For low crowd densities $\eta \leq \eta_c = 0.3 p/m^2$, every single pedestrian walks freely, with an average speed $v_M = 1.5 m/s$ [3], slightly varying from one walker to the other. Conversely, when the crowd density $\eta$ is higher than $\eta_c$, the walking velocity decreases in order to avoid collisions between pedestrians. For very high crowd densities ($\eta \geq \eta_M = 1,6-1,8 \ pedestrians/m^2$ [3]), pedestrians stop walking.

**Simplified footbridge modeling**

The lateral motion of a footbridge can be approximately represented by an Euler-Bernoulli beam equation with viscous damping [2]:

$$m_s(x)\frac{\partial^2 u}{\partial t^2}(x,t) + c(x)\frac{\partial u}{\partial t}(x,t) + k(x)\frac{\partial^4 u}{\partial x^4}(x,t) = F_l(x,t) - m_p(x,t)\frac{\partial^2 u}{\partial t^2}(x,t) \quad (2)$$

where $x$ is the coordinate along the beam axis; $t$ the time; $u(x,t)$ the lateral displacement; $m_s(x)$ the mass per unit length of the beam [$kg/m$]; $c(x)$ the viscous damping coefficient [$N.s/m^2$]; $k(x)$ the stiffness per unit length [$Nm^2$]; $F_l(x,t)$ the pedestrian lateral force per unit length [$N/m$] and $m_p(x,t)$ the linear mass of pedestrians [$kg/m$]. Under the assumption of doubly hinged beam of length $L$, the first lateral mode shape can be approximated by a sinus having the half-period equal to $L$, which is exact when $m_s$, $m_p$, $c$ and $k$ are constant along the beam axis.

Since this mode plays a major role in the lateral footbridge dynamics, the solution is assumed of the form $u(x,t) = U(t)\psi_1(x) = U(t)\sin(\pi x/L)$ and Eq. (2) becomes

$$m^*(t)\ddot{U}(t) + c^*\dot{U}(t) + k^*U(t) = F_l^*(t) \quad (3)$$

where $m^* = m_s^*(t) + m_p^*(t) = \int_0^L [m_s(x) + m_p(x,t)]\psi_1^2 \, dx, \, c^* = \int_0^L c(x)\psi_1^2(x)dx,$

$k^* = \int_0^L k(x)\frac{d^2\psi_1}{dx^2}(x)dx,$ and $F_l^*(t) = \int F_l(x,t)\psi_1(x)dx$. One can observe that the mass $m^*(t)$ is

made of the classical structural mass contribution plus the total mass of pedestrians walking on the footbridge deck and having the instantaneous distribution given by $m_p(x,t)$. Hence, the instantaneous modal frequency of the system footbridge+pedestrians reads:

$$f_{ps}(t) = \frac{1}{2\pi}\sqrt{k^*/m^*(t)} = f_s\left(1 + \frac{2}{L}\int_0^L \frac{m_p(x,t)}{m_s}\sin^2\left(\frac{\pi x}{L}\right)dx\right)^{-1/2} \quad (4)$$

The mass $m_p(x,t)$ and the force $F_l(x,t)$ must be defined to solve Eq. (3) and are related to the approach used for modeling the crowd. Two different approaches are understudy: (i) an "Eulerian" approach consisting in a macroscopic modeling of the crowd which is considered as a whole and (ii) a "Lagrangian" approach consisting in a microscopic modeling of the crowd where each pedestrian is modeled. Under after, only the first approach is presented.

**The Eulerian crowd model (ECM)**

An Eulerian crowd model (ECM) postulates that the crowd behaves like a compressible fluid [3]. This kind of analysis is intended to represent the pedestrian behaviour for high crowd densities. The crowd motion is characterized by it local density $\eta(x,t)$ and its local speed $v(x,t)$. In general three equations govern the motion of a fluid, i.e. the mass conservation

$$\frac{\partial \eta}{\partial t} + \frac{\partial}{\partial x}(\eta.v) = 0 \quad (5)$$

the dynamic equilibrium and a constitutive law. However, for traffic flow modelling, it is usual to substitute these last two equations with a simpler "closure" equation [3, 6], relying on $\eta$ and $v$. For the pedestrian flow, the following closure equation is adopted [3]:

$$v(\eta,\dot{u}) = g(\eta) \times h(\dot{u}) \tag{6}$$

with $g(\eta) = \begin{cases} v_M & \eta \leq \eta_c \\ v_M(1 - \dfrac{1-exp(-\beta(\eta-\eta_c)/(\eta_M-\eta_c))}{1-exp(-\beta)}) & \eta > \eta_c \end{cases}$, $h(\dot{u}) = \begin{cases} 1 - \dfrac{|\dot{u}|}{\dot{u}_{max}} & |\dot{u}| < \dot{u}_{max} \\ 0 & |\dot{u}| \geq \dot{u}_{max} \end{cases}$

The crowd velocity $v(\eta,\dot{u})$ depends on the $g(\eta)$ function, representing the interaction between pedestrians described above. $\beta(x,t)$ is a parameter describing the traffic conditions. The mechanical analogy suggests for $\beta$ the role of a viscosity parameter of the equivalent crowd fluid. A larger $\beta$ value indicates more difficult traffic conditions (highly viscous crowd fluid). Fig. 2 presents the crowd speed on a rigid floor ($h(\dot{u})=1$) for different $\beta$ values. Experimental data lead to the values of $\beta$ being between 0 and 10 [3].

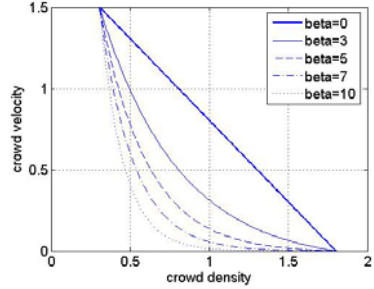

Figure 2- Walking speed vs. crowd density for different $\beta$ values

The second function $h(\dot{u}) \leq 1$ allows accounting for the influence of the lateral footbridge vibration on the crowd speed, in the sense that it imposes a speed reduction when pedestrians walk on a footbridge undergoing large oscillations. In addition, it is assumed that after a stop occurring when $|\dot{u}| \geq \dot{u}_{max}$, the pedestrian speed remains zero during five seconds. After this delay, they begin to walk if $|\dot{u}| < \dot{u}_{max}$, otherwise, the stop lasts five more seconds. Observe that Eq. (6) governs the crowd speed but nothing is said about the pedestrians-structure synchronization phenomenon which also involves the phases of pedestrians and of the structure.

Once the density $\eta(x,t)$ is known, the linear mass density follows $m_p(x,t) = m_{1p}\eta(x,t)\ell(x)$, where $\ell(x)$ is the deck width and $m_{1p}$ is the mass of a single pedestrian. The linear force density reads:

$$F_l(x,t) = f_l(t)\big[\eta(x,t)l(x)\, S(\dot{u},\eta)\big] = f_l(t)\big[n_p(x,t)\, S(\dot{u},\eta)\big] = f_l(t)n_{p,eq}(x,t) \tag{7}$$

where $S(\dot{u},\eta) \in [0;1]$ is a coefficient introduced to represent the synchronization effects [3]. According to the assumptions of Venuti [3], the special case $S=1$ occurs when all the pedestrians walk with the frequency and phase of the structural velocity $\dot{U}(t)$. And Eq.7 states then that the linear force density is given by the product of the force of a single pedestrian with the number $n_p(x,t)$ of pedestrians per unit length. This case is the most severe for the structure, in the sense that the same number of pedestrians always produces a less important structural motion when they are not in this situation of full synchronization.

Conversely, if pedestrians are not synchronized, the total force is defined as the product of the force of a single pedestrian and an equivalent synchronized pedestrian number. This equivalent number is defined to give a *fictitious* total force $F_l(x,t)$ acting at the modal frequency and phase of the structure whose effects (structural motion) are the same as those of the *true* total force, deriving from the true non-synchronized pedestrians (see [2]). The synchronization coefficient, i.e. the ratio between the number of synchronised pedestrians per unit length and the number of pedestrians per unit length, is defined by Venuti [3] as follows

$$S(x,t) = Spp(\eta(x,t)) + Sps(\dot{u}(x,t)) \qquad (8)$$

The first term accounts for the fact that for high densities pedestrians are synchronized each other: for $\eta \leq \eta_c$, the walk is free, every pedestrian has a different frequency and/or phase and the contribution to the total force is therefore null. The second term derives from the observation that for large oscillations, pedestrians have a stronger tendency to be synchronized with the structure. The function $Sps(\dot{u})$ has two branches (Figure 3-b): the first one is a quadratic approximation of the ARUP data [1]. It defines an increase of the synchronization when the structural oscillations are larger. However, after reaching a maximum for a particular amplitude of structural oscillations, the synchronization (and the equivalent pedestrian number) is assumed then to decrease, as represented by the decreasing branch.

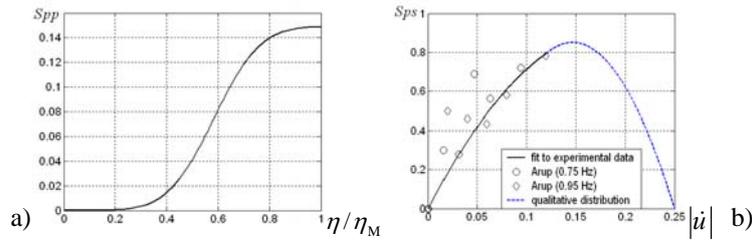

Figure3- a) Coefficient $Spp(\eta)$, b) Coefficient $Sps(\dot{u})$

The previous definition of the synchronisation coefficient (Eq. 8) have some drawbacks: (i) the sum of two contributions is rather "artificial", (ii) the $Spp$ coefficient is arbitrary and not based on experimental data, (iii) it is difficult to distinguish the effects in the case of a vibrating floor between the two synchronisations: the one between pedestrians and the other with the vibrations of the structure ; the synchronisation between pedestrians being induced by the synchronisation with the deck's vibrations. Hence, a new definition of the synchronisation coefficient is introduced. According to SETRA [2], the lateral force induced by a crowd of $N$ pedestrians on a rigid floor can be approximated by the force induced by $N_{eq}$ pedestrians walking in place and having the same phase and frequency. Thus, denoting $\xi$ the proportional modal damping ratio, the synchronisation coefficient in the case of non-vibrating floor is defined as follows

$$Sp(\eta(x,t)) = N_{eq}(x,t)/N = \begin{cases} 8.6\sqrt{\xi/(\eta(x,t)\ell(x)L)} & \text{if } \eta(x,t) \leq \eta_c \\ 1.75/\sqrt{(\eta(x,t)\ell(x)L)} & \text{otherwise} \end{cases}$$

Finally, a new definition for the synchronisation coefficient is given by

$$S(x,t) = \begin{cases} Sp(\eta(x,t)) & \text{if } \ddot{u}(x,t) \leq a_{min} = 0.1 m/s^2 \\ Sps(\dot{u}(x,t)) & \text{otherwise} \end{cases} \qquad (9)$$

**Numerical example : the Millennium Bridge**

Eq. (4) is solved by a Runge-Kutta scheme of order 4-5, and Eq. (5) is discretized by a finite difference scheme (Lax-Friedrichs), fulfilling the Lax Friedrichs conditions for the ratio between the spatial step *dx* and the time step *dt*. The model is validated using the Millennium Footbridge main deck's parameters. The pedestrian mass is assumed to be $75 kg$, hence $m_p(x,t) = 75\eta(x,t)l$. Moreover the following parameters are adopted: $\beta(x,t) = 0$, $U(0) = \dot{U}(0) = 0$ and $\eta(x,0) = 1.2$. Figures 4-a and 4-b represent $\dot{U}(t)$ and $F_l^*(t)$ respectively. The initial high density of pedestrians induces a synchronisation between pedestrians which leads to a lateral force of high amplitudes. As a consequence, the structure's lateral vibrations are more important. When the vibrations' velocity is high ($\dot{u}(x,t) > \dot{u}_{max}$), the pedestrians stop, hence the lateral force amplitude decreases and so do the structural vibrations.

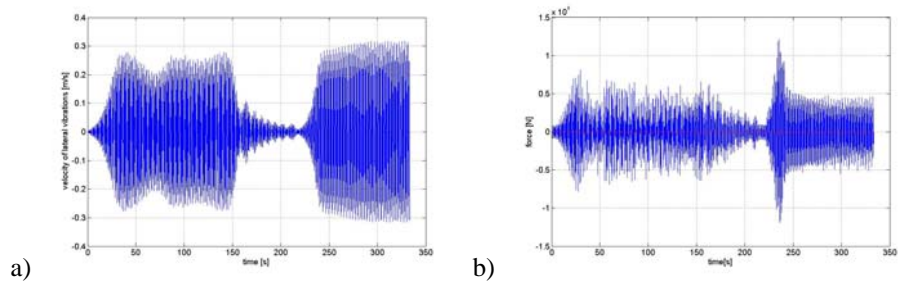

a)  b)

Figure 4 -a) Footbridge velocity $\dot{U}(t)$ ; b) total lateral force $F_l^*(t)$ induced by the crowd

**Conclusion**

In this paper, the crowd flow has been described by an "Eulerian" approach, in a 1D framework, viz. pedestrians are supposed to walk along straight trajectories parallel to the longitudinal dimension of the footbridge. The non-stationary behaviour has been analyzed, as well as the synchronization phenomenon. Some improvements have been proposed. Work is in progress on a "Lagrangian" model including synchronization and the 2D generalization of both approaches.